\documentclass[prl,twocolumn,aps,superscriptaddress,showpacs,floatfix]{revtex4}
\usepackage{amsmath}
\usepackage{graphicx}
\begin{document}
\title{Generation of tunable Terahertz radiation using Josephson vortices}
\author{Sergey Savel'ev}
\affiliation{Frontier Research System, The Institute of Physical
and Chemical Research (RIKEN), Wako-shi, Saitama, 351-0198, Japan}
\author{Valery Yampol'skii}
\affiliation{Frontier Research System, The Institute of Physical
and Chemical Research (RIKEN), Wako-shi, Saitama, 351-0198, Japan}
\affiliation{
Usikov Institute for Radiophysics and
Electronics, Ukrainian Academy of Science, 12 Proskura Street,
61085 Kharkov, Ukraine}
\author{Alexander Rakhmanov}
\affiliation{Frontier Research System, The Institute of Physical
and Chemical Research (RIKEN), Wako-shi, Saitama, 351-0198, Japan}
\affiliation{Institute for Theoretical and Applied Electrodynamics
RAS, 125412 Moscow, Russia}
\author{Franco Nori}
\affiliation{Frontier Research System, The Institute of Physical
and Chemical Research (RIKEN), Wako-shi, Saitama, 351-0198, Japan}
\affiliation{Center for Theoretical Physics, Department of
Physics, University of Michigan, Ann Arbor, MI 48109-1120, USA}
\date{\today}
\begin{abstract}
We propose how to control the THz radiation generated by fast
moving Josephson vortices in spatially modulated (either along the
$c$-axis or the $ab$-plane) samples of $\rm
Bi_2Sr_2CaCu_2O_{8+\delta}$ and related superconducting compounds.
We show that the JVs moving in a subset of weaker junctions can
generate {\it out-of-{\rm ab}-plane} and {\it outside-the-cone}
Cherenkov radiation. The ab-plane modulation of superconducting
properties (achieved, for instance, by ion irradiation
lithography) can result in transition radiation within certain
frequency windows, i.e., allowing the design of tunable THz
emitters and THz photonic crystals.

\end{abstract}
 \maketitle

The recent growing interest in terahertz (THz) science and
technology is due to its many important applications in physics,
astronomy, chemistry, biology, and medicine, including THz
imaging, spectroscopy, tomography, medical diagnosis, health
monitoring, environmental control, as well as chemical and
biological identification \cite{tera-appl}.
The THz gap, that is still hardly reachable for
both electronic and optical devices, covers temperatures of
biological processes and a substantial fraction of the luminosity
remanent from the Big Bang \cite{tera-appl}.

High-temperature $\rm Bi_2Sr_2CaCu_2O_{8+\delta}$ superconductors
have a layered structure that allows
the propagation of electromagnetic waves (called Josephson
plasma oscillations
\cite{plasma,plasma-theor,plasma-exp,plasma-exp1,plasma-exp2})
with Josephson plasma frequency $\omega_J$. This is drastically
different from the strong damping of electromagnetic waves in low
temperature superconductors. It has been recognized (see, e.g.,
\cite{bis-tera,tera2,tachiki}) that the Josephson plasma frequency
lies in THz range.
 A possible way to generate THz radiation in $\rm
Bi_2Sr_2CaCu_2O_{8+\delta}$ and related compounds is to apply an
in-plane magnetic field $H_{ab}$ and an external current
$J_{\parallel c}$ perpendicular to the superconducting layers
(i.e., along the $c$ axis). Josephson vortices (JVs) induced by
$H_{ab}$ and driven fast by the $c$-axis current emit THz
radiation (e.g., \cite{bis-tera,tachiki}). However, it was shown
\cite{mints-kras,cher1,cher2} that the radiation propagates {\it
only\/} along the plane of motion of the JVs and decays in the
$c$-direction. This strong confinement of THz radiation, its
rather restricted controllability, and the large required $c$-axis
current, all limit potential applications.

Recent developments in sample fabrication can produce
superconductors with alternating layers \cite{two-dif-layer}, as
well as spatially-modulated both pinning (e.g., \cite{kwok}) and
superconducting coupling between layers. We propose employing
these capabilities to create modulated structures for a new
generation of THz emitters based on moving JVs.

{\it Radiation from a weak junction:\/} A subset of weaker
intrinsic Josephson junctions in $\rm
Bi_2Sr_2CaCu_2O_{8+\delta}$-- based samples can be made using
either (i) the controllable intercalation technique
\cite{two-dif-layer}, (ii) Chemical Vapor Deposition (CVD) (see,
e.g., \cite{CVD}), or (iii) via the admixture of $\rm
Bi_2Sr_2Cu_2O_{6+\delta}$ and $\rm Bi_2Sr_2Ca_2Cu_3O_{10+\delta}$
\cite{admixture}. The motivation for making such artificial
samples (Fig.~1b) is that the radiation produced by the usual $\rm
Bi_2Sr_2CaCu_2O_{8+\delta}$ materials is confined to the ab-plane.
In other words, if all planes (i.e., all junctions) are identical,
the Cherenkov radiation generated by fast moving JVs
\cite{bis-tera,cher1,cher2,mints-kras} propagates only along the
direction of the JV motion. Electromagnetic (EM) waves along the
$c$-direction cannot be generated in identical-planes samples
because the maximum velocity of the JVs is smaller than, or of the
order of, the smallest velocity of the $c$-axis EM waves. Thus,
previous proposals for pseudo-Cherenkov radiation
\cite{bis-tera,cher1,cher2,mints-kras} in standard $\rm
Bi_2Sr_2CaCu_2O_{8+\delta}$ samples do not refer to the usual
Cherenkov radiation of a fast relativistic particle because the
radiation generated by those JVs is narrowly-confined, i.e., it
does not propagate within a Cherenkov cone.
We propose using an artificial stack of thin $\rm
Bi_2Sr_2CaCu_2O_{8+\delta}$ sheets having several Josephson
junctions which are weaker than the other intrinsic ones. {\it The
JVs in the weaker junctions can move much faster than the
Josephson plasma waves in $\rm Bi_2Sr_2CaCu_2O_{8+\delta}$
materials, producing electromagnetic waves propagating both
parallel and perpendicular to the CuO$_2$ layers} (Fig.~1a,b).
Changing the $c$-axis current changes both the JV speed and the
radiation power.

\begin{figure}[!htp]
\begin{center}
\begin{center}
\includegraphics*[width=8.7cm]{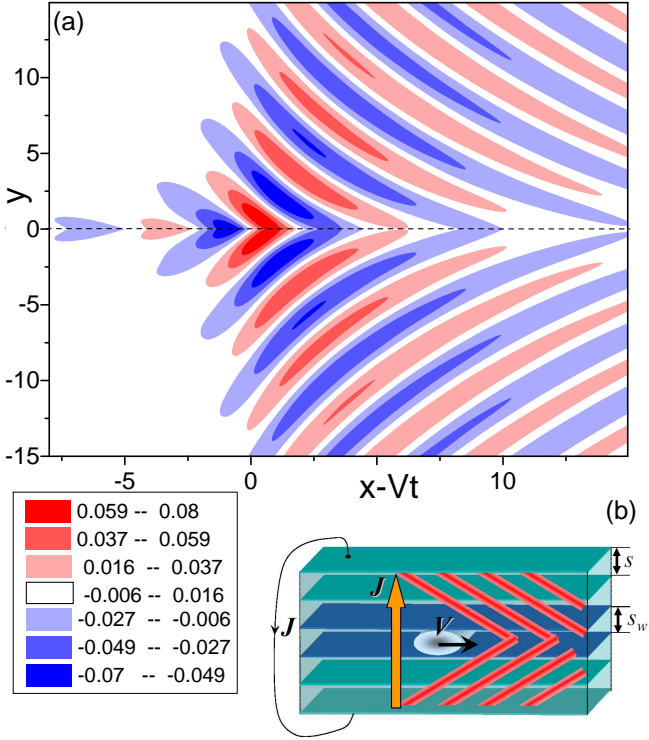}
\end{center}
\vspace{-1.5cm} \caption{Cherenkov radiation generated by a fast
Josephson vortex (located at $x=Vt$) moving in a weaker junction.
(a) Magnetic field distribution $H(x-Vt,y)$ in units of
$\Phi_0/2\pi\lambda_c\lambda_{ab}$ for $J_c^w/J_c=0.2$,
$s_w\epsilon/(s\epsilon_w)=1.2$, $V/V_{\max}^w=0.9$. The
``running'' coordinate, $x-Vt$, is measured in units of $\gamma s
/ (\pi\sqrt{v^2\beta^2-1})$, while the out-of-plane coordinate $y$
is normalized by $s / (\pi\sqrt{v^2\beta^2-1})$, where $\beta=\pi
J_c\epsilon s_w/2J_c^w\epsilon^ws$. The moving vortex emits
radiation propagating forward. This radiation forms a cone
determined by the vortex velocity $V$. (b) A suggested
experimental set up: in a weaker junction an out-of-plane current
$J_{\parallel c}$ drives a Josephson vortex with velocity $V$,
which is higher than the minimum velocity $c_{\min}$ of the
propagating electromagnetic waves.}
\end{center}
\end{figure}

{\it Transition radiation:\/}
A completely different design involves samples with periodically
modulated \cite{ustinov-tr} (in the ab-plane) superconducting
properties which are now uniform along the $c$-axis. For example,
spatial (in-plane) variations of the Josephson maximum $c$-axis
current $J_c$ can be obtained by using irradiation of a standard
$\rm Bi_2Sr_2CaCu_2O_{8+\delta}$ sample covered by a modulated
mask (see, e.g., \cite{kwok}). For such a sample, the Josephson
vortices can produce transition Terahertz radiation (Fig. 2a) even
if the vortex velocity is slower than $c_{\min}$.
The terminology ``transition radiation'' \cite{landau} refers to
the radiation produced when the vortex alternatively transits from
one ``layered medium'' to another.
The spatial periodicity in the $ab$-plane controls the radiation
frequency.

{\it Josephson vortex as a fast relativistic particle.---} Besides
potential applications to THz technology, the motion of a JV in
layered superconducting materials has significant scientific
interest because it can mimic some properties of fast relativistic
particles. Indeed, it is well-known that the Sine-Gordon equation,
describing the motion of a JV in a conventional Josephson
junction, is invariant under a Lorentz transform, where the speed
of light is replaced by the Swihart velocity \cite{barone}. For a
standard Josephson junction, the Swihart velocity restricts both
the maximum speed of small magnetic field perturbations and the
maximum vortex velocity. In layered structures, the equation
describing the Josephson vortex and its magnetic field $H(x,y,t)$
becomes nonlocal \cite{gur}
\begin{equation}\label{h}
H-\lambda_{ab}^2\frac{\partial^2 H}{\partial
y^2}-\lambda_c^2\frac{\partial^2 H}{\partial x^2}+\omega_J^{-2}
\frac{\partial^2}{\partial
t^2}\left(1-\lambda_{ab}^2\frac{\partial^2}{\partial
y^2}\right)H=0,
\end{equation}
\begin{equation}\label{phi}
 \omega_c^{-2}\frac{\partial^2\phi}{\partial
t^2}+\sin\phi\ =\ \frac{l}{\pi}\int_{-\infty}^{\infty}
\mathrm{d}\zeta\;
K_0\!\left(\frac{|\zeta-x|}{\lambda_c}\right)\frac{\partial^2\phi}{\partial\zeta^2}.
\end{equation}
\begin{equation}\label{phi-h}
\frac{\partial \phi}{\partial x}\ =\
\frac{2\pi\lambda_{ab}^2}{\Phi_0}\left\{\frac{\partial
H(x,y=+0)}{\partial y}-\frac{\partial H(x,y=-0)}{\partial
y}\right\}
\end{equation}
Here, the Josephson plasma frequency $\omega_J$ is $\sim$ 1 THz,
depending on doping and temperature; the two length scales
$\lambda_c=\gamma\lambda_{ab}$ and
$\lambda_{ab}=2000/\sqrt{1-T^2/T_c^2}\;$\AA\,, with $T_c=90$K and
$\gamma\sim 300$, determine the characteristic scales of magnetic
field variations parallel (along the $x$-axis) and perpendicular
(along the $y$-axis) to the superconducting layers (Fig.~1b).
Eq.~(\ref{phi}) describes the propagation of the gauge invariant
phase $\phi=\chi_1-\chi_2+2\pi A_y {\tilde s }/\Phi_0$ along the
junction where the Josephson vortex moves, where $\chi_1$ and
$\chi_2$ are the phases of the wave functions of the
superconducting condensate of the CuO$_2$ layers forming the
junction, and $A_y$ is the $y$-component of the vector potential.
Here, $\tilde s$ denotes the thickness of this junction and
$\Phi_0$ is the magnetic flux quantum. Eq.~(\ref{phi}) has its own
space and time scales: $l$ and $\omega_c^{-1}$. There $K_0$
denotes the modified Bessel function usually employed for the
magnetic field distribution in superconductors. When all Josephson
junctions are the same (e.g., as in standard $\rm
Bi_2Sr_2CaCu_2O_{8+\delta}$ samples), $\omega_c=\omega_J$,
${\tilde s}=s$, and the size $l$ of the soliton at $V=0$ is
defined as a product of the anisotropy parameter $\gamma$ and the
spacing $s$ between CuO$_2$ planes ($l\approx \gamma s$). For a
weaker Josephson junction, the Josephson soliton is more
elongated, $l=l_w=\gamma s J_c/J_c^w$, and $\omega_c=\omega_w$ is
determined by the parameters of the weaker junction:
$\omega_w=\omega_J\sqrt{s_wJ_c^w\varepsilon/sJ_c\varepsilon_w}$
with $J_c$, $\varepsilon$ are the maximum superconducting current
and dielectric constant of the intrinsic junctions; the index $w$
refers to the weaker Josephson junction.

 Eq.~(\ref{h}) for the magnetic field $H$ allows us to
obtain the spectrum of EM waves propagating in the layered
structure
\begin{equation}\label{spec}
\omega^2=\omega_J^2+c_J^2(k_y)k_x^2, \ \
c_J(k_y)=\frac{\omega_J\lambda_c}{\sqrt{1+\lambda_{ab}^2k_y^2}}.
\end{equation}
The minimum vortex velocity $c_{\min}$ needed for Cherenkov
radiation ($c_{\min}=c_J(k_y=\pi/s)=\gamma s\omega_J/\pi$), and
the characteristic angle
\begin{equation}
\theta=\arctan\left(\frac{\pi\sqrt{V^2-c_{\min}^2}}{\omega_Js}\right)
\end{equation}
of the propagating EM wave are determined by three standard
conditions: (i) Eq.~(\ref{spec}), (ii) $\omega=k_xV$, and (iii)
the minimum wavelength $k_y\lesssim \pi/s$.
However, the JVs cannot move above the maximum speed $V_{\max}$
determined by the nonlocal Eq.~(\ref{phi}):
$V_{\max}\approx\omega_J\gamma s/2$ for identical-junctions sample
and $V_{\max}^w\approx\omega_w l_w$ for samples with some weaker
junctions. Therefore, the radiation propagating within the
Cherenkov cone with $\theta\ne 0$ cannot be generated using
standard $\rm Bi_2Sr_2CaCu_2O_{8+\delta}$ samples, while it can be
achieved using the proposed artificial stack or $c$-axis-modulated
$\rm Bi_2Sr_2CaCu_2O_{8+\delta}$ samples.
Using Eqs.~(\ref{h}, \ref{phi}) we calculate perturbatively the
Cherenkov radiation (see Fig. 1a):
\[
H_{\rm
Cher}(x,y,t)=\frac{\Phi_0}{2\pi\lambda_{ab}^2}\int_{q_{\min}}^\infty\frac{\mathrm{d}q}{k(q)}\exp[-ql(V)]
\]
\begin{equation}\label{37}
\times\sin[k_x(x-Vt)+k_y(k_x)|y|]\; ,
\end{equation}
where the dynamical-soliton size $l(V)$ decays when the vortex
velocity $V$ \cite{current} increases as:
$l(V)=l_0\left(1+\sqrt{1-V^2/(V_{\max}^{w})^{2}}\right)\!/2$. The
$y$-component $k_y$ of the wave vector is determined by the
dispersion relation (\ref{spec}), while
$q_{\min}=\omega_J/\sqrt{V^2-c_{\min}^2}$. Due to the rather
unusual dispersion relation (\ref{spec}), i.e., the decrease of
$k_y(k_x,\omega=k_xV)$ with increasing $k_x$, the electromagnetic
waves are located {\it outside} the Cherenkov cone (Fig.~1a),
which is drastically different from the Cherenkov radiation of a
fast relativistic particle. The new type of radiation predicted
here could be called {\it outside-the-cone} Cherenkov radiation.

{\it Zone structure of the transition radiation.---} Here we
consider standard $\rm Bi_2Sr_2CaCu_2O_{8+\delta}$ materials with
identical junctions. As mentioned above, previously proposed
pseudo-Cherenkov-like radiation
\cite{bis-tera,cher1,cher2,mints-kras} is confined to the
ab-plane, and requires a  rather strong $c$-axis current. In
contrast to this, we consider an alternative approach to generate
THz radiation using standard $\rm Bi_2Sr_2CaCu_2O_{8+\delta}$
materials which allows to control the spectrum of the radiation.
This proposal might be the easiest to experimentally implement,
among the ones listed here, since it does not require samples with
weaker junctions.

The modulation of superconducting properties along the ab-plane
could be done by using, e.g., either lithography or by controlling
the JV dynamics via their interactions with other type of vortices
(pancake vortices; with its density determined by the out-of-plane
field $H_c$). Indeed, the $c$-axis field, $H_c$, creates another
type of vortices, called pancake vortex stacks, which have slow
dynamics and attract Josephson vortices \cite{cross,cross1}. The
interaction of these two types of vortices changes the dynamics of
the JVs \cite{NM,hirata} and affects their motion in a similar
manner as spatially modulating the critical-current with a period
which can be easily tuned by the out-of-plane field $H_c$.

\begin{figure}[!htp]
\begin{center}
\begin{center}
\includegraphics*[width=8.7cm]{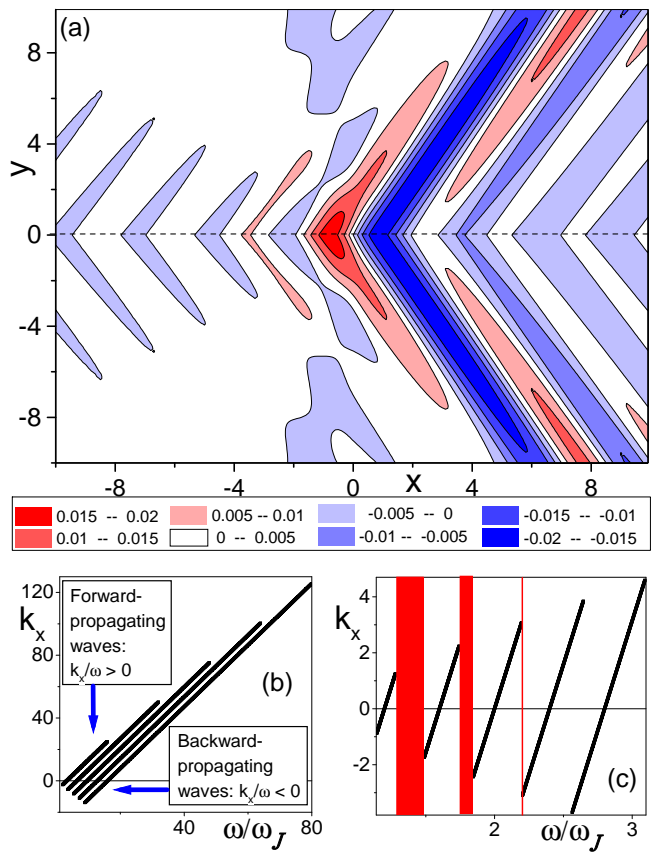}
\end{center}
\vspace*{1cm} \caption{Transition radiation emitted by a Josephson
vortex (located at $x\,=\,0$) moving through a spatially modulated
(along the ab-plane) layered superconducting sample. (a) Magnetic
field distribution $H(x,y)$ (at a certain time moment, say $t=0$)
in units of $\mu_1\Phi_0s/2\pi\lambda_{ab}^2a_{{\rm var}}$ for
$V/c_{\min}=0.8$, $a_{\rm var}/l(V=0)=1$. The in-plane and
out-of-plane coordinates $x$ and $y$ are normalized by the core
size $l$ of a static Josephson vortex and $2s$, respectively.
(b,c) The $x$-component $k_x$ of the wave-vector of the radiation
versus frequency. In contrast to the Cherenkov radiation (Fig.~1),
the phase velocity $k_x/\omega$ of the transition radiation could
be positive or negative (b), resulting in waves propagating both
forward and backward with respect to the Josephson vortex motion.
The radiation frequency has forbidden zones, shown by red strips
in (c), when the vortex moves relatively slow. This suggests the
remarkable possibility of THz photonic crystals in modulated
layered superconductors, which is potentially important for
applications.}
\end{center}
\end{figure}

To simplify the problem, we study the radiation of Josephson
vortices with modulated $\omega_c(x)$. Note that the modulation of
any other parameters (e.g., $l(x)$ or both $l(x)$ and
$\omega_c(x)$) in Eq.~(\ref{phi}) results in the same qualitative
results. For periodically-varying properties (with spatial period
$a_{\rm var}$) of intrinsic junctions,  the frequency $\omega$ and
the in-plane wave vector $k_x$ of the emitted EM waves are related
by the condition $\omega=(k_x+2\pi m/a_{\rm var})V$, with $m=1, 2,
\dots$. This is the momentum conservation law: the momentum of a
moving JV can be transferred not only to the EM waves but also to
the modulated medium, in full analogy with electron motion in a
periodic potential. Following this analogy, we derived the ``zone
structure'' of the radiated waves
$H=\sum_m\int_{\omega_{\min}(m)}^{\omega_{\max}(m)}H_m(\omega)\sin(k_x
x+k_y(\omega,k_x) |y| - \omega t)\;d\omega$, where
\begin{equation} \label{aa}
H_m(\omega)=\frac{\mu_m\Phi_0}{2\pi\lambda_{ab}^2a_{\rm var}}
\;\frac{\omega k_x(m)}{k_y(\omega,k_x(m))}\;
\frac{\exp\left(-|\omega|l(V)/V\right)}{|k_x(m)|\;V_{\max}\omega_J-\omega^2}.
\end{equation}
Here $k_x(m)$ is related to the momentum $2\pi m/a_{\rm var}$
transferred to the modulated medium by $k_x=\omega/V-2\pi m/a_{\rm
var}$ and $\mu_m$ is the Fourier component of
$2[1-\omega_c(x)/\omega_J]$. For a chosen zone-number $m$, the
radiation is emitted in a certain frequency window
$\omega_{\min}(m)<\omega<\omega_{\max}(m)$, with
\[
\omega_{\min,\,\max}(m)=\frac{2\pi m\; c_{\min}}{a_{\rm var
}}\;\frac{V^2}{c_{\min}^2-V^2}
\]
\vspace{.1cm}
\begin{equation}\label{50}
\times\left[\frac{c_{\min}}{V}\mp
\sqrt{1-\omega_J^2\left(\frac{a_{\rm var}}{2\pi
mc_{\min}}\right)^2\frac{(c_{\min})^2-V^2}{V^2}}\,\right].
\end{equation}
Since the frequencies $\omega_{\min,\,\max}$ change with changing
$a_{\rm var}$, the generation of THz radiation in a certain
frequency range could be designed by adjusting the period of the
spatial modulations of ion irradiation during the sample
preparation. When the out-of-plane magnetic field $H_c$ is
applied, the transition THz radiation might be tuned by varying
$H_c$.

In contrast to the outside-the-cone Cherenkov-like radiation
(Fig.~1a), our proposed transition radiation (Fig.~2a) propagates
both {\it forward and backward} in space. Indeed, the in-plane
component $k_x$ of the wave vector and, thus, the corresponding
phase velocity both change sign (Fig.~2b,c). Also, the EM waves
running backward can be directly seen from the magnetic field
distribution shown in Fig.~2a.
For fast Josephson vortices moving with velocity close to
$c_{\min}$, the frequency zones overlap for different zone number
$m$ (Fig.~2b).
 However, for slower speeds, we
obtain the forbidden frequency ranges
$\omega_{\max}(m)<\omega<\omega_{\min}(m+1)$ (see, Fig. 2c) of
radiated electromagnetic waves. Thus, such materials might be used
for making tunable THz photonic crystals controlled, e.g., by the
changing out-of-plane magnetic field $H_c$.

{\it Conclusions.---} A grand challenge is to controllably
generate electromagnetic waves in $\rm Bi_2Sr_2CaCu_2O_{8+\delta}$
and other layered superconducting compounds because of its
Terahertz frequency range \cite{bis-tera,tera2,tachiki}.
Considering recent advances in sample fabrication
\cite{two-dif-layer,kwok}, we propose several experimentally
realizable systems for generating continuous radiation in a
controllable frequency range.

We gratefully acknowledge conversations with M. Tachiki, A.V.
Ustinov, M. Gaifullin and partial support from the NSA and ARDA
under AFOSR contract No. F49620-02-1-0334, and by the NSF grant
No. EIA-0130383.

\end{document}